\documentclass[sigconf]{acmart}
\usepackage{enumitem}
\usepackage{multirow}

\renewcommand\footnotetextcopyrightpermission[1]{}
\AtBeginDocument{%
  \providecommand\BibTeX{{%
    \normalfont B\kern-0.5em{\scshape i\kern-0.25em b}\kern-0.8em\TeX}}}

\setcopyright{acmlicensed}
\copyrightyear{2018}
\acmYear{2018}
\acmDOI{XXXXXXX.XXXXXXX}

\acmConference[Conference acronym 'XX]{Make sure to enter the correct
  conference title from your rights confirmation emai}{June 03--05,
  2018}{Woodstock, NY}
\acmISBN{978-1-4503-XXXX-X/18/06}

\begin{document}

% \title{CELA: Cost-Efficient Language Model Alignment for CTR Prediction}
\title{CELA: Cost-Efficient Language Model Alignment for CTR Prediction}
% \title{Cost-Efficient Language Model Alignment for CTR Prediction}

\author{Xingmei Wang}
\affiliation{
  \country{University of Science and Technology of China, China}
  % \city{Hefei}
   % \country{China}
}
\email{xingmeiwang@mail.ustc.edu.cn}

\author{Weiwen Liu}
\affiliation{
  \country{Huawei Noah's Ark Lab, China}
  % \city{Shenzhen}
  % \country{China}
}
\email{liuweiwen8@huawei.com}

\author{Xiaolong Chen}
\affiliation{
  \country{University of Science and Technology of China, China}
  % \city{Hefei}
  % \country{China}
}
\email{chenxiaolong@mail.ustc.edu.cn}

\author{Qi Liu}
\author{Xu Huang}
\affiliation{
  \country{University of Science and Technology of China, China}
  % \city{Hefei}
  % \country{China}
}
\email{qiliu67@mail.ustc.edu.cn}
\email{xuhuangcs@mail.ustc.edu.cn}

% \author{Xu Huang}
% \affiliation{
%   \country{University of Science and Technology of China, China}
%   % \city{Hefei}
%   % \country{China}
% }
\email{xuhuangcs@mail.ustc.edu.cn}

\author{Yichao Wang}
\author{Xiangyang Li}
\affiliation{
  \country{Huawei Noah's Ark Lab, China}
  % \city{Shenzhen}
  % \country{China}
}
\email{wangyichao5@huawei.com}
\email{lixiangyang34@huawei.com}

% \author{Yasheng Wang}
% \affiliation{
%   \country{Huawei Noah's Ark Lab, China}
%   % \city{Shenzhen}
%   % \country{China}
% }
% \email{wangyasheng@huawei.com}

\author{Yasheng Wang}
\author{Zhenhua Dong}
\affiliation{
  \country{Huawei Noah's Ark Lab, China}
  % \city{Shenzhen}
  % \country{China}
}
\email{wangyasheng@huawei.com}
\email{dongzhenhua@huawei.com}

\author{Defu Lian}
\affiliation{
  \country{University of Science and Technology of China, China}
  % \city{Hefei}
  % \country{China}
}
\email{liandefu@ustc.edu.cn}
\authornote{The corresponding author.}

\author{Ruiming Tang}
\affiliation{
  \country{Huawei Noah's Ark Lab, China}
  % \city{Shenzhen}
  % \country{China}
}
\email{tangruiming@huawei.com}

\newcommand{\lxy}[1]{{\color{green} [lxy: #1]}}
\newcommand{\ww}[1]{{\color{orange} [ww: #1]}}
\newcommand{\yc}[1]{{\color{purple} [yichao: #1]}}
\renewcommand{\shortauthors}{Wang and Liu, et al.}

\begin{abstract}
  Click-Through Rate (CTR) prediction holds a paramount position in recommender systems. The prevailing ID-based paradigm underperforms in cold-start scenarios due to the skewed distribution of feature frequency. Additionally, the utilization of a single modality fails to exploit the knowledge contained within textual features.
Recent efforts have sought to mitigate these challenges by integrating Pre-trained Language Models (PLMs). They design hard prompts to structure raw features into text for each interaction and then apply PLMs for text processing. With external knowledge and reasoning capabilities, PLMs extract valuable information even in cases of sparse interactions. Nevertheless, compared to ID-based models, pure text modeling degrades the efficacy of collaborative filtering, as well as feature scalability and efficiency during both training and inference.
To address these issues, we propose \textbf{C}ost-\textbf{E}fficient \textbf{L}anguage Model \textbf{A}lignment (\textbf{CELA}) for CTR prediction. CELA incorporates item textual features and language models while preserving the collaborative filtering capabilities of ID-based models. This model-agnostic framework can be equipped with plug-and-play textual features, with item-level alignment enhancing the utilization of external information while maintaining training and inference efficiency. Through extensive offline experiments, CELA demonstrates superior performance compared to state-of-the-art methods. Furthermore, an online A/B test conducted on an industrial advertising recommender system showcases its practical effectiveness, solidifying the potential for real-world applications of CELA. Codes are available at \href{https://github.com/pepsi2222/CELA}{https://github.com/pepsi2222/CELA}.
\end{abstract}


% \ccsdesc[500]{Information systems~Recommender systems; Language models}
% \ccsdesc[500]{Computing methodologies~Neural networks}




\maketitle

\section{Introduction}
Recommender systems deliver personalized items to users by analyzing user browsing history and demographics, presenting users with items tailored to their preferences. Click-through Rate (CTR) prediction plays a crucial role in this process by estimating the likelihood that the user will interact with a specific item. This ensures that the displayed items are those most likely to bolster user engagement and revenue.

\begin{figure}[t]
    \centering
    \includegraphics[width=1\linewidth]{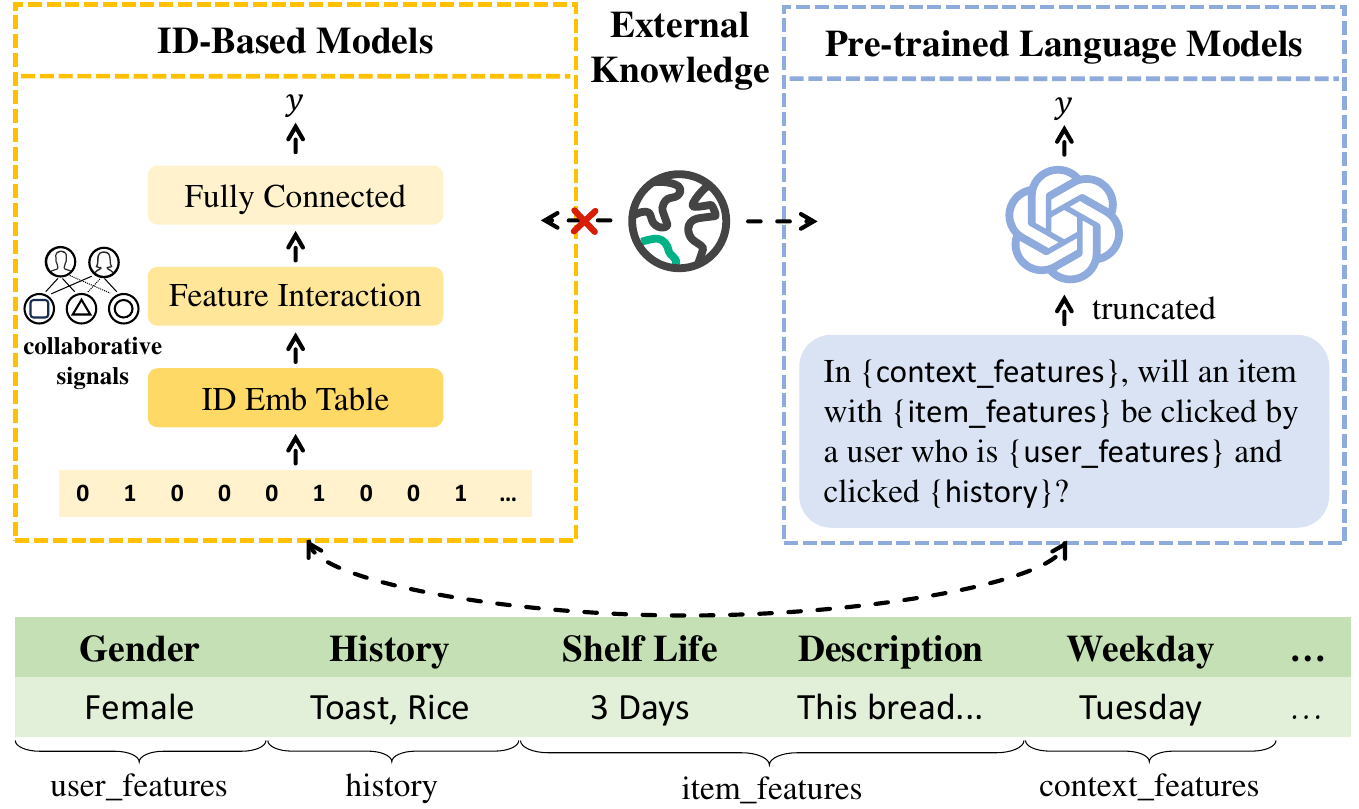}
    % \vspace{-0.2cm}
    \caption{ID-Based Models vs. Pre-trained Language Models (PLMs) for CTR prediction. The former converts tabular data into one-hot vectors, capturing collaborative signals without external knowledge. The latter populates data into a hard prompt, and then feeds truncated texts into the PLM, leveraging external world knowledge for data interpretation.}
    \label{fig:example}
\end{figure}

The ID-based paradigm has garnered considerable success. Notable examples include DeepFM~\cite{guo2017deepfm}, PNN~\cite{qu2016product}, and DCN~\cite{wang2017deep}, which focus on high-order explicit feature crossing, while DIN~\cite{zhou2018deep} and DIEN~\cite{zhou2019deep} are geared towards mining behavior sequences. This paradigm, illustrated on the left side of Figure~\ref{fig:example}, entails encoding user and item features into sparse one-hot vectors, subsequently transforming into dense embedding vectors via a lookup process. Latent co-occurrences among features are discerned via a feature interaction layer. Over the past decade, the advancement of these models has centered on developing more sophisticated and scenario-adapted feature interaction layers as a means to enhance overall performance.
Despite their prevalence, ID-based models exhibit certain limitations: 
1) Dependence on historical data: This dependence can result in suboptimal modeling in scenarios such as cold-start or long-tail elements, where historical interactions for items or features are limited or absent, leading to inadequate learning of embeddings.
2) Limited feature scope: Primarily focusing on categorical and numerical features, ID-based models overlook other modalities like text, which may harbor untapped knowledge that could potentially enhance the model's effectiveness.

To alleviate these limitations, works such as PTab~\cite{liu2022ptab} and CTRL~\cite{li2023ctrl}
% Rella~\cite{lin2023rella}, and TallRec~\cite{bao2023tallrec} 
have integrated Pre-trained Language Models (PLMs) into recommender systems. These models employ hard prompts to structure raw features for each interaction, subsequently processed through PLMs for recommendation purposes, as shown on the right side of Figure~\ref{fig:example}. By adopting natural language to represent features instead of one-hot vectors, this approach preserves the semantic integrity of features. This enables PLMs to apply their external knowledge and reasoning capabilities, effectively capturing valuable information even in cases of sparse interactions. For example, ID-based models struggle to learn embeddings for infrequently interacted items like bread, whereas PLMs utilize external knowledge to classify rice as a type of food, akin to toast. Concurrently, this natural language representation facilitates the integration of textual features, such as item descriptions, thereby augmenting the recommender system's effectiveness. 

However, these text-enhanced models face inherent limitations that hinder their practical deployment, shown in Figure~\ref{fig:example}. 
% Firstly, the token-wise approach in NLP task modeling might not be optimal granularity for capturing feature co-occurrences, often leading to diminished performance. 
First, modeling based exclusively on pure text may not be ideally suited for capturing collaborative signals, which often results in reduced performance~\cite{li2023ctrl}. 
% This limitation is likely attributable to the tokenization process disrupting feature-wise interactions\lxy{add reference}.
Second, scalability issues emerge as the number of features increases, potentially leading to constructed text exceeding the language model's maximum token limit and thus losing valuable information, particularly when the sequence of user behaviors is extensively long. Furthermore, it becomes impractical to delineate user behaviors in a detailed manner, including the specification of item features for each behavior~\cite{yang2024item}.
Last but not least, to bridge the gap between PLM outputs and recommendation tasks, PLMs are usually fine-tuned at the interaction-level~\cite{liu2022ptab,geng2022recommendation,li2023ctrl,muhamed2021ctr}, yet this process is notably time-intensive. Such inefficiency inherent in PLM text processing extends to inference as well~\cite{liu2022ptab,geng2022recommendation,lin2023rella,bao2023tallrec}.
% language models generally exhibit lower efficiency in text processing compared to ID-based models in tabular data handling. This inefficiency is not restricted to training~\cite{liu2022ptab,geng2022recommendation,li2023ctrl,muhamed2021ctr} but extends to the inference as well~\cite{liu2022ptab,geng2022recommendation,lin2023rella,bao2023tallrec}.

To overcome these challenges, we propose the \textbf{C}ost-\textbf{E}fficient \textbf{L}anguage Model \textbf{A}lignment (\textbf{CELA}) for CTR prediction. 
% CELA leverages item textual features and PLMs to mitigate the cold-start issue and expand the feature scope. It integrates ID-based models to preserve the collaborative filtering effectiveness, and shifts away from the hard prompt mode, thereby addressing scalability concerns. Through item-level alignment, the PLM is fine-tuned effectively and oriented towards recommendation tasks.
To mitigate cold-start issues, CELA expands its feature scope by leveraging item text features and  PLMs to generate more accurate item and user profiles. To maintain optimality and reduce inference latency, CELA integrates ID-based models, sustaining the efficacy and efficiency of collaborative filtering. For scalability, CELA shifts from the hard prompt mode to using PLMs solely as encoders for item textual features. Furthermore, to reduce training costs, CELA fine-tunes PLMs through item-level alignment.
CELA is structured as a three-phase paradigm:
1) Domain-Adaptive Pre-training stage, the PLM is further pre-trained on a corpus of item textual features, tailoring it to the specific dataset; 
2) Recommendation-Oriented Modal Alignment stage, contrastive learning is employed to align the PLM with the item ID embedding table of a developed ID-based model. This alignment is at the item level, ensuring minimal training overhead;
and 3) Multi-Modal Feature Fusion stage, aligned text representations of each item are cached, facilitating efficient access via item identifiers, thereby reducing inference costs. Then these text representations are integrated into a new ID-based model, along with non-textual features, enabling their interplay to identify potential click patterns. 
The final two stages are structured for alternate execution, ensuring ongoing alignment of the PLM with the progressively refined embedding table.

The main contributions of this paper are summarized as follows:
\begin{itemize}[left=0pt]
    \item We propose a novel paradigm for integrating item textual features with ID-based models. This approach is model-agnostic and necessitates minimal modifications to the existing network architecture, rendering it conveniently applicable.
    \item We develop an item-level alignment strategy to direct the capabilities of the PLM for recommendation tasks, thereby reducing training time overhead and achieving performance enhancements. Moreover, the cached item text representations are indexed by item identifiers during inference, ensuring low latency.
    \item Comprehensive offline experiments on public and industrial datasets demonstrate the effectiveness of CELA. An online A/B test within %Huawei AppGallery 
    an industrial advertising recommender system further validates CELA's real-world efficacy and feasibility. % achieving a 0.97\% increase in eCPM and a 0.92\% rise in Download-Through Rate (DTR).
\end{itemize}

\section{Related Works}
\subsection{ID-Based Models for CTR Prediction}

ID-based models have been predominant in industry recommendations over the past decade.
They capture the characteristics of the items and users by one-hot vectors and learn the collaborative signals to predict the likelihood of a user clicking on an item~\cite{lian2020personalized}. 
Their architectures are meticulously crafted, evolving into increasingly intricate ones. 
FM~\cite{rendle2010factorization} stands out as one of the earliest machine learning models to incorporate second-order feature interactions. 
Both WideDeep~\cite{cheng2016wide} and DeepFM~\cite{guo2017deepfm} incorporate deep components to learn high-level feature interactions. PNN~\cite{qu2016product} extends the feature interaction function and tightly couples the interaction of first-order and second-order features.
% CCPM~\cite{liu2015convolutional} and FGCNN~\cite{liu2019feature} employ convolutional kernels to capture local patterns and hierarchical spatial relationships within the ID embeddings.
AutoInt~\cite{song2019autoint} and DESTINE~\cite{xu2021disentangled} incorporate self-attention mechanisms to weight different feature interactions.
% Fi-GNN~\cite{li2019fi} leverages graph structures to model feature interactions.
% AFM~\cite{xiao2017attentional}, HFM~\cite{tay2019holographic} and LorentzFM~\cite{xu2020learning} explores a more diverse set of second-order feature interaction functions.
DCN~\cite{wang2017deep}, DCNv2~\cite{wang2021dcn}, and EDCN~\cite{chen2021enhancing} capture meaningful feature crossings automatically, eliminating the need for manual feature engineering or exhaustive searching.
DIN~\cite{zhou2018deep} and DIEN~\cite{zhou2019deep} model user behavior sequences, aiming to capture the relationships between the target item and historical behaviors.
% Researchers are continuing to explore ways to improve CTR models.
% However, pure ID-based models miss out on capturing semantic relationships and the uniqueness of the IDs prevents it from being shared between different realms.

\subsection{Text-Enhanced Models for CTR Prediction}

Language models have made remarkable strides, achieving unprecedented natural language understanding and generation. 
% These models are pre-trained on massive datasets and adopt advanced architectures like Transformer~\cite{vaswani2017attention}, enabling diverse applications such as chatbots~\cite{du2022glm}, translation~\cite{zhang2022opt}, and content creation~\cite{chung2022talebrush}.
Pre-trained on extensive datasets and utilizing sophisticated architectures like Transformer~\cite{vaswani2017attention}, these models support diverse applications such as chatbots~\cite{du2022glm}, translation~\cite{zhang2022opt}, and content creation~\cite{chung2022talebrush}.
Highlighting their potential in semantic comprehension, language models are considered promising for recommendation. Analyzing extra-textual elements such as product descriptions, these models may have the capability to discern user intent and item nuances, thereby enhancing ID-based ones.
% P5~\cite{geng2022recommendation} represents a paradigm that redefines various tasks into natural language processing (NLP) tasks. By structuring interactions and metadata in natural language and inputting them into T5~\cite{raffel2020exploring}, it can generate the target response.
P5~\cite{geng2022recommendation} redefines tasks as natural language processing (NLP) tasks, structuring interactions and metadata in natural language and inputting them into T5~\cite{raffel2020exploring} to generate target responses.
% CTR-BERT~\cite{muhamed2021ctr} concatenates all features along with their respective feature names. It undergoes pre-training and distillation with BERT~\cite{devlin2018bert}. Subsequently, BERT encodes the text of users and items, generating representations that serve as inputs for a late fusion MLP.
CTR-BERT~\cite{muhamed2021ctr} concatenates all features with their names, undergoes pre-training and distillation with BERT~\cite{devlin2018bert}, and encodes user and item text to generate representations for a late fusion MLP.
% PTab~\cite{liu2022ptab} is pre-trained on BERT with a language head using the "field: value" corpus, which is derived from tabular data. It is then fine-tuned for the CTR task, incorporating a classifier head.
PTab~\cite{liu2022ptab} is pre-trained on BERT using a "field: value" corpus from tabular data, then fine-tuned for the CTR task with a classifier head.
UniSRec~\cite{hou2022towards} employs a frozen BERT as a text encoder and utilizes a Transformer architecture to model user behaviors.
CTRL~\cite{li2023ctrl} uses contrastive learning to align representations between collaborative and semantic models at the interaction level. The collaborative model, now enriched with semantic information, is fine-tuned and can be deployed independently.
% Despite advancements, text-enhanced models are less favored in scenarios like e-commerce due to three reasons: 1) suboptimal performances~\cite{geng2022recommendation}\cite{liu2022ptab} compared to well-established ID-based counterparts; 2) high inference costs~\cite{geng2022recommendation}\cite{liu2022ptab} causing latency issues in real-time applications; and 3) elevated training expenses~\cite{geng2022recommendation}\cite{li2023ctrl}\cite{liu2022ptab}\cite{muhamed2021ctr}, making them less economically viable than alternative solutions.

\section{Preliminary}
\subsection{Task Formulation}
CTR models are designed to estimate the probability that a user will engage with a specific item—such as music, news, or advertisement—by clicking on it. Mathematically, a CTR model $f$ fits the likelihood $P(\cdot)$ of a click event $y=1$ based on a variety of input features $\boldsymbol{x}$, such as user profile, browsing history, item characteristics, and contextual information. 
% such as user profile $\boldsymbol{x}^u$, browsing history $\boldsymbol{x}^h$, item characteristics $\boldsymbol{x}^i$, and contextual information $\boldsymbol{x}^c$. 
The predicted CTR is formulated as:
\begin{equation}
    \small
    \hat{y}=f(\boldsymbol{x}).
\end{equation}

Consider a dataset $\mathcal{D} = \{(\boldsymbol{x}_k, y_k) | k = 1, \ldots, N\}$, with $N$ as the total number of interactions. Each pair $(\boldsymbol{x}_k, y_k)$, where $y_k \in \{0, 1\}$, represents the $k$-th pair of input features and its corresponding ground truth label. Typically, the model $f$ is trained by minimizing the binary cross-entropy loss:
\begin{equation}
\small
    \mathcal{L}_{BCE}=
    -\frac{1}{N}
    \sum^{N}_{k=1}
        \left[
            y_k\log(\hat{y}_k)+(1-y_k)\log(1-\hat{y}_k)
        \right].
    \label{eq:bce}
\end{equation}

\subsection{Embedding Layer}
Raw input features are usually categorical and then converted into one-hot or multi-hot vectors. 
In one-hot encoding, each category is represented by a unique binary vector, whose length matches its cardinality $C$, and only one bit is set to 1 to indicate the the presence of that category,
as shown in Figure~\ref{fig:example}.
In contrast, multi-hot encoding allows for presence of multiple categories at once. 
% For instance, raw features comprising Gender={\ttfamily Female}, History={\ttfamily items 0,1}, Shelf life={\ttfamily 3 Days}, Weekday={\ttfamily Tuesday}, can be represented as:
% \begin{equation*}
%     \begin{aligned}
%         \boldsymbol{x}&=
%             \boldsymbol{x}^u \circ \boldsymbol{x}^h \circ \boldsymbol{x}^i \circ \boldsymbol{x}^c \\ 
%             &=\underbrace{[1, 0]}_\text{\ttfamily Gender}
%             \circ
%             \underbrace{[1,1,0,\ldots,0]}_\text{\ttfamily History}
%             \circ
%             \underbrace{[0,0,1,\ldots,0]}_\text{\ttfamily Shelf life}
%             \circ
%             \underbrace{[0,1,0,\ldots,0]}_\text{\ttfamily Weekday}
%     \end{aligned},
% \end{equation*}
% where $\circ$ denotes the concatenation function.

To compress those sparse, high-dimensional vectors into dense, lower-dimensional ones, the embedding layer is utilized. 
It acts as a lookup table, denoted by $\mathbf{E}^{(\cdot)}\in \mathbb{R}^{C^{(\cdot)} \times d}$ , which maps each categorical feature into a dense vector of a fixed size $d$. Here, $d$ represents the embedding size. 
Let $\boldsymbol{x}^u$ denote user profile, $\boldsymbol{x}^h$ denote browsing history, $\boldsymbol{x}^c$ denote contextual information, $\boldsymbol{x}^{i,n}$ denote non-textual features of candidate item $i$ and $\boldsymbol{x}^{h,n}$ denote non-textual features in user history.
This process can be formalized as:
\begin{equation}
\boldsymbol{e}=\boldsymbol{x}^u \mathbf{E}^u \circ \boldsymbol{x}^{h,n} \mathbf{E}^h \circ \boldsymbol{x}^{i,n} \mathbf{E}^i\circ \boldsymbol{x}^c \mathbf{E}^c,
\end{equation}
where $\circ$ denotes the concatenation function.

\section{Methodology}
\begin{figure*}[tb]
    \centering
    \includegraphics[width=0.9\textwidth]{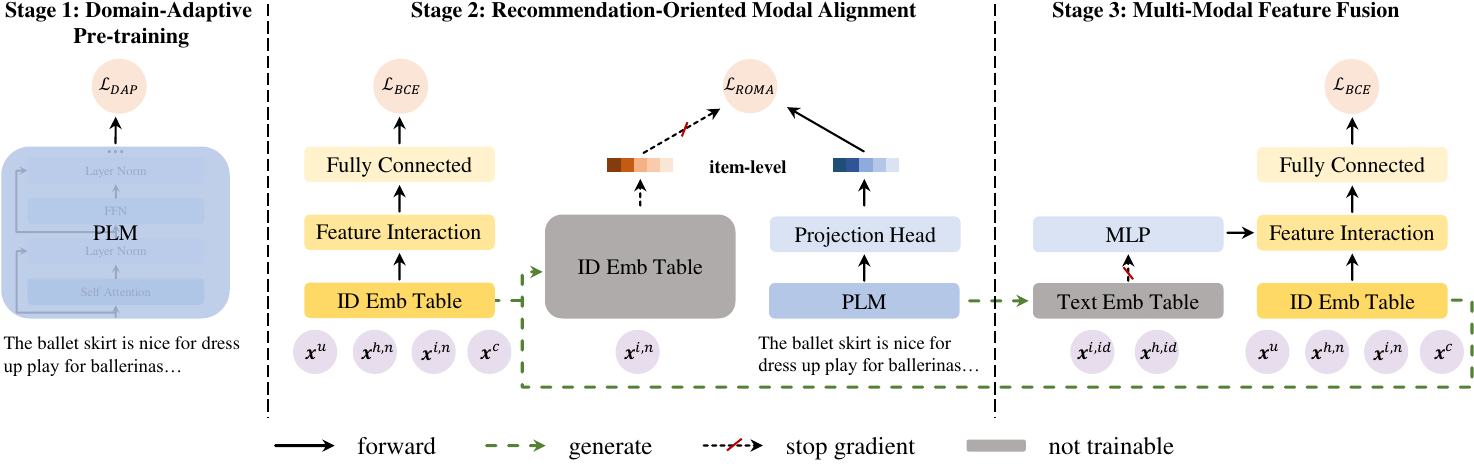}
    \caption{The overall framework of CELA. The first stage pre-trains a PLM on domain-specific item texts. In the second stage, an ID-based model is developed and item text representations from the PLM are aligned with the item-side feature embeddings of the ID-based model in latent space. The third stage merges aligned text representations with non-textual features for training a new ID-based model. The final two stages are executed alternately, as denoted by the green dotted line.}
    \label{fig:overview}
\end{figure*}

The proposed framework is depicted in Figure~\ref{fig:overview}, encompassing three stages:
\begin{itemize}[left=0pt]
    \item \textbf{Domain-Adapted Pre-training (DAP).} It involves further pre-training of a PLM on a corpus comprising item-side text specific to the recommendation dataset, to better adapt the PLM to the unique characteristics of the dataset. 
    % Concurrently, a traditional ID-based model is trained.
    \item \textbf{Recommendation-Oriented Modal Alignment (ROMA).} It focuses on aligning the text representations of items from the PLM with the item-side feature embeddings from the ID-based model in the latent space. Such alignment is essential for tailoring the text representations to recommendation tasks.
    \item \textbf{Multi-Modal Feature Fusion (MF$^2$).} The aligned item text representations are treated as a common feature, which is then integrated with non-textual feature embeddings. Subsequently, a new ID-based model is trained to leverage both the aligned text representations and other features, thereby enhancing the overall effectiveness of the recommender system.
\end{itemize}

Notably, the last two stages can be executed alternatively, where the PLM is continually aligned to the more advanced ID embedding table.

\subsection{DAP: Domain-Adaptive Pre-training}

To enhance the PLM's understanding of dataset-specific texts, we construct a corpus from item-side texts, rich in detailed item descriptions, as illustrated in Appendix~\ref{appendix:corpus}. Leveraging this corpus for further pre-training, we utilize the Masked Language Model Loss (MLM)~\cite{devlin2018bert}, which masks a portion of tokens within a sentence and calculates the loss for each masked token prediction. For instance, the input text "This ballet skirt is nice for dress-up play for ballerinas...," may be masked as "This ballet skirt is [MASK] for dress-up play for [MASK]...", forcing the PLM to predict the mask tokens and give the MLM loss as:
% \begin{equation}
%     \boldsymbol{h}=PLM(\boldsymbol{x}^{i,text})
% \end{equation}
% \begin{equation}
%     \boldsymbol{p}_m = LMHead(\boldsymbol{h})
% \end{equation}
% \begin{equation}
%     \mathcal{L}_{MLM}=-\sum_m
%     CrossEntropy(\boldsymbol{p}_m, \boldsymbol{x}^{i,text}_m)
% \end{equation}
\begin{equation}
\small
    \mathcal{L}_{MLM}=-
    \sum_{i\in\mathcal{I}}
    \sum_m
    \log P
    \left(
        \boldsymbol{x}^{i,t}_m|\boldsymbol{x}^{i,t}_{\verb|\|m}
    \right),
\end{equation}
where $\mathcal{I}$ denotes the item space, $\boldsymbol{x}^{i,t}$ denotes item $i$'s textual feature, $m$ and $\backslash m$ denote the masked tokens and the rest of the tokens, respectively.

Simultaneously, to enhance item text representations and mitigate the anisotropy issue~\cite{ethayarajh2019contextual,li2020sentence}—where learned representations are constrained to a narrow cone in vector space, reducing expressiveness—we incorporate SimCSE~\cite{gao2021simcse} into our framework. 
SimCSE, an unsupervised technique, employs contrastive learning, treating a text as its positive pair by feeding it into the PLM twice with standard dropout as the only noise. By drawing positive pairs closer, the PLM is compelled to discern the meanings of texts amidst noise, thereby improving its comprehensive capacity. Moreover, distancing the representations of distinct texts promotes a more uniform distribution of representations in vector space, addressing the anisotropy and yielding superior text representations.
This process can be formulated as:
\begin{equation}
\small
    \mathcal{L}_{SimCSE}=-
    \sum_{i\in\mathcal{I}}
    \log\frac{
    \exp{
        \left(
            sim\left(
                \boldsymbol{h}_i, \boldsymbol{h}'_i
            \right)/\tau_1
        \right)
        }
    }{
        \sum_{j\in\mathcal{I}}
        \exp
        \left(
            sim\left(
                \boldsymbol{h}_i, \boldsymbol{h}'_j
            \right)/\tau_1
        \right)
    },
\end{equation}
where $sim(\cdot)$ denotes the similarity function, $\tau_1$ denotes the temperature scaling the similarity measure, and $\boldsymbol{h}_i$ and $\boldsymbol{h}'_i$ denote the representations of item $i$'s text.

The composite loss function for the PLM's pre-training is:
\begin{equation}
    \mathcal{L}_{DAP} = \mathcal{L}_{MLM} + \alpha \mathcal{L}_{SimCSE},
\end{equation}
where $\alpha$ is a weighting parameter that balances the contributions of the MLM loss and SimCSE loss. This integrated approach ensures a comprehensive and nuanced adaptation of the PLM to the specifics of the dataset.

\subsection{ROMA: Recommendation-Oriented Modal Alignment}
Despite the profound domain understanding that the PLM achieves through pretraining, a significant gap persists between their output space and that of collaborative filtering. This discrepancy arises because the capabilities of the PLM extend well beyond what recommendation tasks require. Without targeted training and constraints, the outputs of the PLM may not align with the needs of recommender systems. Consequently, it is crucial to refine and orient the PLM's outputs towards recommendation tasks, a process referred to as alignment. This essential alignment, which aims to provide a representation of the item's textual feature that can be seamlessly integrated with other features for the next stage, is illustrated in the central portion of Figure~\ref{fig:overview}. The substantial number of parameters in the PLM and the extensive volume of interactions in the dataset render interaction-level alignment inefficient~\cite{li2023ctrl,liu2022ptab,geng2022recommendation}. Consequently, we propose item-level alignment as a solution.
% Despite a refined representation of item texts is obtained, a notable discrepancy exists between this representational space and that of the collaborative filtering space. To harmonize the representation of item text with other features, a critical process of cross-modal alignment is undertaken at this stage. This alignment, aimed at reconciling the disparities between the two representational spaces, is exemplified in the central portion of Figure~\ref{fig:overview}. 

\subsubsection{ID-Based Model}
\label{old_id_based_model}
To provide a reference for steering capabilities of the PLM towards recommendation tasks, we develop an ID-based model that processes exclusively categorical or numerical features. The model's prediction function is formulated as:
\begin{equation}
    \hat{y}=f
            \left(
                \boldsymbol{x}^u, 
                \boldsymbol{x}^{h,n}, 
                \boldsymbol{x}^{i,n}, 
                \boldsymbol{x}^c
            \right).
\end{equation}
The corresponding training loss is defined in Equation~\ref{eq:bce}.

\subsubsection{Item-Level Modal Alignment}
First, the textual and non-textual features on the item side are encoded into their respective representations. For item $i$, its non-textual features $\boldsymbol{x}^{i, n}$, are encoded utilizing the frozen ID embedding table developed in Section~\ref{old_id_based_model}. Conversely, the textual feature $\boldsymbol{x}^{i, t}$ is processed through the PLM, followed by a transformation via a projection head. This head is to adjust the dimensionality of the output from the PLM to match that of the non-textual features. This encoding process is formulated as:
\begin{equation}
\boldsymbol{e}^{n}_i = \boldsymbol{x}^{i, n}\mathbf{E}^i,
\end{equation}
\begin{equation}
\boldsymbol{e}^{t}_i = PLM(\boldsymbol{x}^{i, t}) \mathbf{W} + \boldsymbol{b},
\end{equation}
where $\boldsymbol{e}^{n}_i$ represents the encoded representation of item $i$'s non-textual features, and $\boldsymbol{e}^{t}_i$ represents the encoded textual feature. The terms $\mathbf{W}$ and $\boldsymbol{b}$ are the parameters of the projection head, representing the matrix and bias, respectively.

Contrastive learning is then utilized to align the representations of textual and non-textual features on the item side. Within a given batch, representations of these features from the same item are considered positive pairs, while unmatched representations are treated as negative pairs. Given that both modalities depict the same item, there is an inherent overlap of insights between its positive pair, analogous to the relationship among visual, textual, and auditory data~\cite{girdhar2023imagebind}. By incentivizing the model to differentiate between matching and mismatched pairs, contrastive learning facilitates the establishment of correspondence or mapping functions across modalities.
The alignment begins with
\begin{equation}
\small
    \mathcal{L}_{t,n}=-\frac{1}{\left|\mathcal{I}\right|}
                                    \sum_{i\in\mathcal{I}}
                                    \log
                                    \frac{
                                        \exp
                                        \left(
                                            sim\left(
                                                \boldsymbol{e}^{t}_i, \boldsymbol{e}^{n}_i
                                            \right)/\tau_2
                                        \right)
                                        }
                                        {
                                            \sum_{k\in\mathcal{I}}
                                            \exp
                                            \left(
                                                sim\left(
                                                    \boldsymbol{e}^{t}_i, \boldsymbol{e}^{n}_k
                                                \right)/\tau_2
                                            \right)
                                        },
\end{equation}
where $\tau_2$ denotes another temperature scaling the similarity.

To maintain symmetry in the alignment process, a corresponding loss for aligning non-textual representations to textual ones is defined as:
\begin{equation}
\small
    \mathcal{L}_{n,t}=-\frac{1}{\left|\mathcal{I}\right|}
                                    \sum_{i\in\mathcal{I}}
                                    \log
                                    \frac{
                                        \exp
                                        \left(
                                            sim\left(
                                                \boldsymbol{e}^{n}_i, \boldsymbol{e}^{t}_i
                                            \right)/\tau_2
                                        \right)
                                        }
                                        {
                                            \sum_{k\in\mathcal{I}}
                                            \exp
                                            \left(
                                                sim\left(
                                                    \boldsymbol{e}^{n}_i, \boldsymbol{e}^{t}_k
                                                \right)/\tau_2
                                            \right)
                                        }.
\end{equation}
The overall modal alignment loss combines these two components:
\begin{equation}
    \mathcal{L}_{ROMA} =\mathcal{L}_{t,n} + \mathcal{L}_{n,t}.
\end{equation}

Training is exclusively confined to the PLM and projection head, with the ID embedding table remaining non-trainable to prevent disruption of the collaborative filtering space. It is important to emphasize that, the item-level alignment significantly reduces the computational resources and time required for training.

% In this stage, the training is exclusively confined to the language model side, while the embedding table remains static and non-trainable. This delineation is critical; training the embedding table would disrupt the integrity of the space occupied by the non-textual features. Specifically, it would no longer maintain its collaborative filtering space, but rather, it would become influenced and potentially biased by the language model. Keeping the embedding table non-trainable ensures that the collaborative filtering space remains unaffected by updates to the language model.

\subsection{MF$^2$: Multi-Modal Feature Fusion}
To mitigate suboptimal performance caused by purely text-based models, we preserve collaborative signals captured by the ID-based model while integrating the aligned PLM. Serving solely as an item textual feature encoder, the PLM avoids scalability issues from increased features and prevents information loss by encoding textual features of user behaviors item by item.

% To minimize both the training duration and the online inference latency, a text embedding table for items is generated. This process involves inputting the text of each item into the aligned PLM to derive a text embedding, which is then stored in the text embedding table. This approach enables efficient access of specific text embedding using the item's identifier, thereby obviating the need for real-time processing through the PLM.
To reduce training time and online latency, we create an item text embedding table. Text for each item is processed through the aligned PLM to obtain the embedding, and stored for quick access by item identifiers, eliminating the need for real-time PLM processing.
Then, as depicted on the right side of Figure~\ref{fig:overview}, the recommender system is enhanced by integrating the item's textual feature alongside non-textual features. This item textual feature, indexed from the text embedding table using the item's identifier, is processed using an MLP to ensure dimensional consistency. Within the feature interactions layer, textual and non-textual features are treated uniformly. Such uniformity across all modalities prevents substantial modifications to the original network, thus facilitating its application to any network architecture.
This process is formulated as:
\begin{equation}
    \boldsymbol{e}' = \boldsymbol{e}
                        \circ 
                        MLP\left(
                            \boldsymbol{x}^{i,id}~\mathbf{E}^{i,t}
                        \right)
                        \circ 
                        MLP\left(
                            \boldsymbol{x}^{h,id}~\mathbf{E}^{i,t}
                        \right)
                        ,
\end{equation}
\begin{equation}
    \hat{y} = FC\left(
                    FI\left(
                        \boldsymbol{e}'
                    \right)
                \right),
\end{equation}
where $\mathbf{E}^{i,t}$ denotes the text embedding table, $\boldsymbol{x}^{i,id}$ denotes the candidate item's ID, and $\boldsymbol{x}^{h,id}$ denotes items' IDs in user history. $\boldsymbol{e}'$ is an augmented version of $\boldsymbol{e}$, concatenated with textual embeddings of the candidate item and user history. $FI(\cdot)$ denotes the feature interaction layer and $FC(\cdot)$ denotes the fully connected layer.

\subsection{Alternate Training}

The last two stages, as indicated by the green dotted line in Figure~\ref{fig:overview}, are designed to be executed alternately. Following the completion of the ID embedding table training in the MF$^2$ stage, we can revert to the ROMA stage. Here, the item text representation is realigned to the newly trained ID embedding table. Subsequently, this realigned PLM is utilized to generate a new text embedding table. This updated table is then employed in the subsequent iteration of the MF$^2$ stage for further training. This alternate approach allows for iterative refinement of both the alignment and fusion processes, thereby progressively improving the overall performance.

\section{Experiments}
In this section, we conduct extensive offline and online experiments to assess CELA by answering these research questions (RQs).
\begin{itemize}[left=0pt]
    \item \textbf{RQ1:} How does CELA perform against the state-of-the-art?
    \item \textbf{RQ2:} How is CELA's efficiency during training and inference?
    \item \textbf{RQ3:} How does CELA perform in cold-start scenarios?
    \item \textbf{RQ4:} How do different components influence CELA?
    % \item \textbf{RQ5:} How do different sizes of language models and various backbones affect the performance?
    % \item \textbf{RQ6:} Does CELA exhibit cross-architecture generalization?
    % \item \textbf{RQ6:} Does alignment of cold-start items hinder performance?
    % \item \textbf{RQ7:} How does CELA fare in real online scenarios?
\end{itemize}

\subsection{Experimental Settings}

\subsubsection{Datasets}
We conduct offline experiments on three datasets, with details provided in Appendix~\ref{appendix:datasets}.

\subsubsection{Baselines}
Our method is benchmarked against contemporary state-of-the-art models, encompassing both ID-based and text-enhanced models.
Classic ID-based models, such as \textbf{AutoInt}~\cite{song2019autoint}, \textbf{DCN}~\cite{wang2017deep}, \textbf{DCNv2}~\cite{wang2021dcn}, \textbf{FiBiNET}~\cite{huang2019fibinet}, \textbf{PNN}~\cite{qu2016product} , \textbf{WideDeep}~\cite{cheng2016wide}, \textbf{DeepFM}~\cite{guo2017deepfm}, and \textbf{xDeepFM}~\cite{lian2018xdeepfm}, focus on modeling high-order feature interactions. In contrast, models like \textbf{DIN}~\cite{zhou2018deep} and \textbf{DIEN}~\cite{zhou2019deep} specialize in fine-grained user behavior modeling. Text-enhanced models, including \textbf{P5}~\cite{geng2022recommendation} and \textbf{PTab}~\cite{liu2022ptab}, aim to describe data in natural language to elicit responses from language models. \textbf{CTR-BERT}~\cite{muhamed2021ctr} integrates traditional and textual features for fusion in an MLP, \textbf{UniSRec}~\cite{hou2022towards} utilizes a frozen BERT~\cite{devlin2018bert} as its text encoder and \textbf{CTRL}~\cite{li2023ctrl} aligns textual and tabular data representations to infuse the ID-based model with semantics.
\textbf{BoW}~\cite{zhang2010understanding} uses bag-of-words representations as item text representations.

\subsubsection{Evaluation Metrics}
In this research, Area Under the Curve (AUC) and logarithmic loss (Logloss) are used as evaluation metrics for CTR models~\cite{song2019autoint,wang2021dcn,qu2016product,guo2017deepfm}. AUC assesses ranking ability, while Logloss measures predictive accuracy.
% In this research, Area Under the Curve (AUC) and logarithmic loss (Logloss) are adopted as evaluation metrics, both widely acknowledged in the assessment of CTR models~\cite{song2019autoint,wang2021dcn,qu2016product,guo2017deepfm}. AUC measures the model's ability to rank positive instances above negative ones, indicating ranking efficacy. Logloss quantifies the correspondence between predicted and actual CTR, reflecting predictive accuracy.
% In this research, Area Under the Curve (AUC) and logarithmic loss (Logloss) are adopted as evaluation metrics, both widely acknowledged in the assessment of CTR models~\cite{song2019autoint,wang2021dcn,qu2016product,guo2017deepfm}. AUC evaluates the model's capacity to accurately rank a positive instance above a negative one, thereby indicating its ranking efficacy. Logloss, on the other hand, quantifies the degree of correspondence between the predicted and actual CTR, offering a metric for the model's predictive accuracy.

\subsection{Overall Performance (RQ1)}

\begin{table}
\vspace{-0.3cm}
\setlength{\tabcolsep}{3.5pt}
  \caption{
  Performance comparison with baselines. The best results are in \textbf{bold}, second-best are in \underline{underlined}. $^\ast$ denotes statistical significance (p < 0.05) compared to the second-best.}
  \label{tab:overall}
  \scalebox{0.9}{
  \begin{tabular}{@{}cccccccc@{}}
\toprule
\multirow{2}{*}{Model}& \multicolumn{2}{c}{Amazon} & \multicolumn{2}{c}{MovieLens} & \multicolumn{2}{c}{AppGallery}\\
\cmidrule(r){2-3} \cmidrule(lr){4-5} \cmidrule(l){6-7}
& AUC & Logloss & AUC & Logloss & AUC & Logloss \\
\midrule
AutoInt         & 0.8793     & 0.4384     & 0.8348        & 0.3694          & 0.8426      & 0.4777                     \\
DCN             & 0.8784     & 0.4407     & 0.8352        & 0.3673          & 0.8420      & 0.4787                     \\
DCNv2           & 0.8809     & 0.4335     & 0.8377        & 0.3659          & 0.8432      & 0.4771                     \\
FiBiNET         & 0.8792     & 0.4365     & 0.8353        & 0.3647          & 0.8428      & 0.4779                     \\
PNN             & 0.8839     & 0.4314     & 0.8361        & 0.3670          & 0.8431      & 0.4774                     \\
WideDeep        & 0.8798     & 0.4383     & 0.8348        & 0.3675          & 0.8422      & 0.4787                     \\
DeepFM          & 0.8798     & 0.4378     & 0.8375        & 0.3642          & 0.8434      & 0.4771                     \\
xDeepFM         & 0.8811     & 0.4332     & 0.8367        & 0.3679          & 0.8435      & 0.4775                     \\
DIN             & 0.8949     & 0.4110     & 0.8367        & 0.3626          & 0.8437      & 0.4771                     \\
DIEN            & 0.8937     & 0.4183     & 0.8343        & 0.3659          & 0.8435      & 0.4816                     \\
\midrule
P5              & 0.8847     & 0.4406     & 0.8197        & 0.3820          & 0.8328      & 0.4920                    \\
PTab            & 0.8811     & 0.4582     & 0.8227        & 0.3809          & 0.8364      & 0.4897                     \\
CTR-BERT        & 0.8763     & 0.4406     & 0.8294        & 0.3676          & 0.8384      & 0.4846                     \\
UniSRec         & 0.8826     & 0.4316     & 0.8381        & \underline{0.3620} & 0.8459   & 0.4736                     \\
CTRL            & 0.8954 & \underline{0.4100} & \underline{0.8385} & 0.3655   & \underline{0.8464} & \underline{0.4729}  \\
BoW             & \underline{0.8965}     & 0.4110   & 0.8371    & 0.3656 & 0.8457 & 0.4729 \\
% ReLLa           & 0.7528	& 0.6442     & 0.6298   & 0.4791    & 0.& \\
\midrule
CELA            & \textbf{0.8996*} & \textbf{0.3996*} & \textbf{0.8426*} & \textbf{0.3574*} & \textbf{0.8481*}   & \textbf{0.4704*}  \\

\bottomrule

\end{tabular}}
\vspace{-0.3cm}
\end{table}

We conduct a comparative performance evaluation of CELA against baselines, with the results detailed in Table~\ref{tab:overall}. We observe that:
\begin{itemize}[left=0pt]
    \item PLMs enhance ID-based models as item textual feature encoders. UniSRec and CELA surpass ID-based models on MovieLens and AppGallery. This demonstrates that PLMs capture valuable information from item textual features not exploited by collaborative filtering, thereby improving model efficacy.
    % \item Within ID-based models, DIN and DIEN excel on Amazon, because their fine-grained modeling of user behaviors caters to the pronounced reliance on user behaviors. Conversely, for Movielens and AppGallery, the dependency on user behaviors appears less critical, resulting in no marked advantage for DIN and DIEN.
    \item Purely text-based models typically underperform ID-based models. This is evidenced by the performance of P5, and PTab, which are inferior to most ID-based models. These findings are consistent with the results reported by \cite{li2023ctrl}, which underscore the limitations of exclusive reliance on semantic modeling.
    % \item Among text-enhanced models, P5 and PTab perform worse than most ID-based models. This observation is consistent with results from \cite{li2023ctrl}, corroborating that recommender systems based purely on text generally underperform compared to ID-based models.
    \item The integration of semantic and collaborative signals leads to improved performance. This is demonstrated by CTRL, UniSrec, and CELA, which excel in most cases. The injection of semantics into the ID-based model distinguishes them from traditional ID-based models. Furthermore, the retention of collaborative signals makes them outperform other text-enhanced models.
    \item Task-oriented alignment is crucial for PLMs to eliminate the representation discrepancy and better adapt to downstream tasks. By aligning the outputs of PLMs and ID-based models, CTRL and CELA outperform UniSRec across three datasets. This outcome indicates that a simple whitening process in UniSRec is insufficient to bridge the gap. Therefore, alignment is essential to tailor PLMs for recommendation purposes.
    \item Using PLMs solely as item textual feature encoders prevents information loss from feature proliferation. Compared to CTRL, CELA performs better in all cases since it extracts textual information from behavior sequences by encoding each item's textual feature individually, thus avoiding lengthy annotations resulting from including all features of an item. In contrast, CTRL transforms tabular data into textual data using a hard prompt template and takes the PLM as a textual data encoder, which fails to provide detailed annotations for items in user behaviors due to token number limitations, leading to information loss.
    % \item CELA demonstrates superior performance across all datasets, thereby validating three core assertions: firstly, the utilization of the PLM extracts intrinsic knowledge within the textual feature; secondly, employing the PLM solely as a textual feature encoder prevents information loss caused by an increase in features and enables the detailed delineation of items; thirdly, the integration of collaborative signals significantly enhances model efficacy. This last point is further supported by the superior outcomes observed in the CTRL and UniSRec compared to other baselines.
\end{itemize}

\subsection{Efficiency Measurement (RQ2)}

\begin{table}
    \caption{Comparison of training and inference efficiency between different models on AppGallery. Params is the number of trainable parameters, the inference time is calculated for one batch and the total training time is for the whole process.}
    \vspace{-0.2cm}
    \tabcolsep=0.4cm
    \label{tab:overhead}
    \scalebox{0.85}{
    \begin{tabular}{@{}crrr@{}}
\toprule
Model       & Params (M)    & Inference (ms)    & Training (h)    \\
\midrule
DIN         & 0.58          & 61.38             & 1.76                  \\
P5          & 60.75         & 1125.13           & 170.68                \\
PTab        & 102.38        & 256.54            & 405.72                \\  % 168.95 + 236.77
CTR-BERT    & 102.98        & 241.55            & 96.41                 \\  % 1.07 + 95.34
UniSRec     & 0.54          & 68.49             & 3.04                  \\
CTRL        & 102.98        & 61.38             & 52.29                 \\  % 50.86 + 1.43
BoW         & 1.07          & 70.48             & 1.59          \\
% ReLLa       & 0             &                   & 0             \\
\midrule
CELA DAP    & 102.88        & /                 & 1.07                  \\
CELA ROMA   & 102.88        & /                 & 2.06                  \\
% CELA DAP    & (102.88+0.58) 103.46        & /          & (1.07+1.76) 2.83      \\
% CELA ROMA   & 102.30                      & /          & 0.30                  \\
CELA MF$^2$ & 0.63          & 69.64             & 2.29                  \\
\midrule
CELA        & /             & 69.64             & 5.42                  \\
\bottomrule
\end{tabular}}
% \vspace{-0.2cm}
\end{table}

To explore the practicality of CELA in industrial scenarios, we measure its training duration and inference latency on the industrial dataset AppGallery. Results presented in Table~\ref{tab:overhead} reveals that:
\begin{itemize}[left=0pt]
    \item Models utilizing hard prompt templates are impractical for real-world applications due to high training costs or inference overhead. This is attributable to their model complexity, indicated by the number of parameters. Training models like P5, PTab, CTR-BERT, and CTRL on extensive interaction-level textual data requires significant time, leading to infrequent updates and the use of outdated models, which negatively impacts performance. Additionally, the high inference overhead of P5, PTab, and CTR-BERT renders them unsuitable for real-time serving.
    % \item Within text-enhanced models, the training expenses of P5, PTab, CTR-BERT, and CTRL are notably high, attributable to their complexity, which is implied by the number of parameters. the inference overhead of P5, PTab, and CTR-BERT is also very high for the same reason. Although reducing the update frequency can reduce training costs, employing an outdated language model adversely affects recommendation performance. Therefore, it is not practical to apply them in real-world scenarios.
    \item UniSRec and CELA maintain training and inference costs comparable to DIN. However, UniSRec's performance is limited due to the lack of fine-tuning of the PLM. This limitation arises from insufficient alignment of the PLM's knowledge with downstream tasks, resulting in suboptimal integration of textual and non-textual features. In contrast, CELA fine-tunes the PLM with negligible training expense, achieving superior performance.
    % \item UniSRec, by not fine-tuning the language model, maintains training and inference costs comparable to those of DIN. However, using a frozen language model limits its efficacy in recommendation tasks. This limitation stems from the insufficient alignment of the PLM's knowledge with the specific requirements of the recommendation domain, resulting in suboptimal integration of textual and non-textual features.
    \item CELA is efficient in both training and inference, making it industry-friendly. Its low training expense results from fine-tuning the PLM at the item level rather than the interaction level, as the number of items is significantly lower than the number of interactions, with a difference of up to 550 times here. Moreover, CELA's inference cost is comparable to that of DIN as it avoids real-time text processing, utilizing an additional text embedding table that occupies only 87.2MB of storage space.
    % \item The ROMA stage of CELA necessitates fine-tuning of the language model, but this tuning is conducted at the item level rather than the interaction level. Typically, the number of items is significantly lower than that of interactions, leading to minimal training costs, particularly in contexts with a limited number of items. The training and inference costs of CELA are on par with those of DIN, making CELA industry-friendly.
\end{itemize}

\subsection{Performance w.r.t. Long-tail Items (RQ3)}

\begin{figure}
    \centering
    \includegraphics[width=0.9\linewidth]{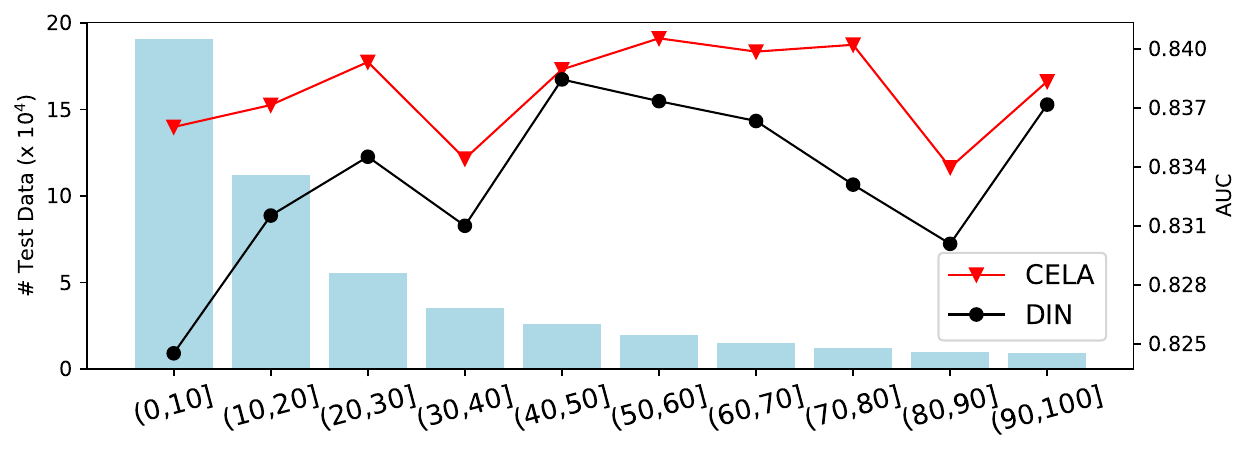}
    \caption{Performance comparison w.r.t. long-tail items on Amazon. The bar graph represents the number of interactions within test data for each group, while the line chart represents the AUC.}
    \label{fig:cold_start}
    \vspace{-0.2cm}
\end{figure}

To assess CELA's efficacy in cold-start scenarios, the test dataset is segmented into groups based on item popularity, with the AUC evaluated for each group. As illustrated in Figure~\ref{fig:cold_start}, CELA consistently outperforms the baseline across all cold-start conditions. 
This superior performance is attributed to the incorporation of a PLM that leverages embedded external knowledge for text comprehension. Notably, when the popularity falls below 20, the improvements of CELA become particularly pronounced. This is because, under such conditions, ID-based models struggle to learn effective ID embeddings owing to their dependence on historical interactions.

\subsection{Ablation Study (RQ4)}
To explore the impact of different training processes and components on the overall performance of CELA, we conduct an ablation study and the results are presented in Table~\ref{tab:ablation}.
Firstly, we propose a variant to verify that the performance gains of CELA are attributed to its semantics rather than additional parameters:
\begin{itemize}[left=0pt]
\item \textbf{Scaled DIN:} Replacing CELA’s non-trainable text embedding table with a trainable one.
\end{itemize}
\subsubsection{Domain-Adpative Pre-training}
Domain-Adpative Pre-training (DAP) improves performance by enhancing the PLM’s understanding and expressiveness of dataset-specific texts. We design three variants to verify its effectiveness:
\begin{itemize}[left=0pt]
    \item \textbf{DAP w/o MLM:} The MLM loss is removed in DAP.
    \item \textbf{DAP w/o SimCSE:} The SimCSE loss is removed in DAP.
    \item \textbf{CELA w/o DAP:} The DAP stage is removed from CELA.
\end{itemize}
As shown in Table~\ref{tab:ablation}, removing either MLM or SimCSE losses, or both, results in performance degradation compared to the full CELA, demonstrating that both losses contribute to performance improvement. However, using only the SimCSE loss in pre-training, without the MLM loss, performs worse than no pre-training on the Amazon dataset. This indicates that SimCSE loss is more suitable as an auxiliary objective to MLM loss, as originally proposed in \cite{gao2021simcse}.

\subsubsection{Recommendation-Oriented Modal Alignment}
Recommendation-Oriented Modal Alignment (ROMA) boosts performance by aligning the outputs of the PLM with the ID-based model at the item level. We designed three variants to verify its effectiveness and explore what accounts for the enhancement:

\begin{itemize}[left=0pt]
\item \textbf{CELA w/o ROMA:} No alignment is applied, and the text embedding table in MF$^2$ is encoded by a non-aligned PLM.
\item \textbf{CELA w/o semantics:} The text embedding table in MF$^2$ is replaced by ROMA's ID embedding table, excluding any semantics.
\item \textbf{ROMA w/o PLM:} The PLM in ROMA is replaced by a pre-encoded text embedding table, with only the table's weights being directly updated.
\end{itemize}

Removing ROMA results in performance degradation, affirming the indispensability of task-oriented alignment of PLM outputs. The performance of CELA w/o semantics is comparable to that of DIN but significantly lags behind full CELA, indicating that CELA’s performance gains stem from the integration of semantics and imply negligible semantic loss during training. Moreover, directly applying ROMA to the text embedding table, instead of the original PLM, leads to reduced performance due to less preservation of semantics, evident from the reduced number of parameters. This underscores the necessity of PLM in ROMA.

\begin{table}
  \caption{Ablation study. The language model is RoBERTa and the backbone is DIN.}
  \vspace{-0.3cm}
  \label{tab:ablation}
  \tabcolsep=0.3cm
  \scalebox{0.9}{
  \begin{tabular}{ccccc}
\toprule
\multirow{2}{*}{Variants}& \multicolumn{2}{c}{Amazon} & \multicolumn{2}{c}{AppGallery} \\
\cmidrule(r){2-3} \cmidrule(l){4-5}
& AUC & Logloss & AUC & Logloss \\
\midrule
DIN                 & 0.8949 & 0.4110 & 0.8437  & 0.4771\\
Scaled DIN	        & 0.8951 & 0.4079 & 0.8452  & 0.4747\\
\midrule
DAP w/o MLM         & 0.8981 & 0.4047 & 0.8477  & 0.4711\\
DAP w/o SimCSE      & 0.8989 & 0.4033 & 0.8478  & 0.4710\\
CELA w/o DAP        & 0.8986 & 0.4012 & 0.8475  & 0.4715\\
\midrule
CELA w/o ROMA       & 0.8965 & 0.4050 & 0.8475  & 0.4709\\
CELA w/o semantics  & 0.8959 & 0.4089 & 0.8440  & 0.4765\\
ROMA w/o PLM        & 0.8669 & 0.5534 & 0.8463  & 0.4731\\
% \midrule
% ROMA w/ cos         & \textbf{0.8997} & 0.4001 & 0.8470 & 0.4715\\
% MF$^2$ w RoBERTa       & 0.8952 & 0.4127 & 0.8472  & 0.4719\\
\midrule
CELA                & \textbf{0.8996} & \textbf{0.3996}  & \textbf{0.8481} & \textbf{0.4705}\\
\bottomrule
\end{tabular}}
\vspace{-0.2cm}
\end{table}

\subsection{Further Analysis}
% \subsection{Compatibility Study (RQ3)}
\subsubsection{Compatibility for language models}
\begin{figure}
    \centering
    \includegraphics[width=1\linewidth]{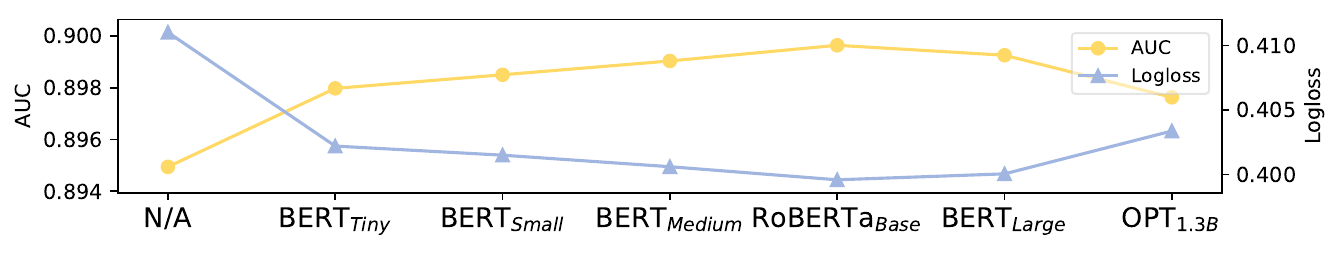}
    \caption{Performance comparison of CELAs with different PLMs on Amazon. The backbone is indicated by N/A.}
    \label{fig:diff_size}
    \vspace{-0.1cm}
\end{figure}

We compare the performance of CELA equipped with PLMs of various sizes: BERT$_{Tiny}$ (14.5M)~\cite{jiao2019tinybert}, BERT$_{Small}$ (29.1M)~\cite{bhargava2021generalization}, BERT$_{Medium}$ (41.7M)~\cite{turc2019well}, RoBERTa$_{Base}$ (110M)~\cite{liu2019roberta}, BERT$_{Large}$ (336M)~\cite{devlin2018bert}, and OPT (1.3B)~\cite{zhang2022opt}. For the OPT model, Causal Language Modeling (CLM) pre-training loss and LoRA fine-tuning~\cite{hu2022lora} are implemented. Experimental results are illustrated in Figure~\ref{fig:diff_size}, leading to the following insights:
\begin{itemize}[left=0pt]
    % \item CELA, when enhanced with a PLM, no matter its size, exhibits significant performance gains over DIN, validating that the use of PLMs can extract knowledge from textual features and benefit CTR prediction.
    \item CELA, enhanced with a PLM regardless of size, significantly outperforms DIN. This validates that PLMs can effectively extract knowledge from item textual features, benefiting CTR prediction.
    % \item As the complexity of the comprehensive PLM escalates, the marginal gains in CELA's performance exhibit a plateauing trend. Notably, BERT$_{Large}$ underperforms relative to RoBERTa$_{Base}$, attributable to the former's significantly larger parameter set, which does not fine-tune as effectively with limited items. This observation implies that beyond a specific threshold, the addition of parameters ceases to yield proportional improvements, suggesting that a medium-sized PLM is sufficient to attain optimal performance.
    \item As the complexity of the comprehensive PLM increases, CELA's performance gains plateau. Notably, BERT$_{Large}$ underperforms compared to RoBERTa$_{Base}$ due to its larger parameter set, which does not fine-tune effectively with limited items. This suggests that beyond a certain threshold, additional parameters do not yield proportional improvements, indicating that a medium-sized PLM is sufficient for optimal performance.
    % \item CELA with OPT exhibits a decrease in performance compared to comprehensive PLMs. This reduction is not only because the extensive parameter space of such models proves challenging to fine-tune effectively with a limited item dataset, but it is also likely due to the generative PLM's unsuitability as a text encoder for CTR prediction.
    \item Generative decoder-based large language models do not enhance CELA as effectively as comprehensive PLMs. Specifically, CELA with OPT shows decreased performance compared to comprehensive PLMs. This reduction is due to the difficulty in fine-tuning large parameter models with limited data and the generative PLM's unsuitability as a text encoder for CTR prediction.
\end{itemize}

% \begin{table}
%     \caption{Performance comparison of models with different sizes of language models. The backbone is DIN.}
%     \label{tab:diff_size}
%     \begin{tabular}{@{}ccc@{}}
% \toprule
%  Language Model & AUC & Logloss \\
% \midrule
% N/A                 & 0.8949 & 0.4110 \\
% TinyBERT (14.5M)    & 0.8980 & 0.4022 \\
% bert-small (29.1M)  & 0.8985 & 0.4015 \\
% bert-medium (41.7M) & 0.8990 & 0.4006 \\
% Roberta (110M)      & \textbf{0.8996} & \textbf{0.3996} \\
% bert-large (336M)   & 0.8993 & 0.4001 \\
% \bottomrule
% \end{tabular}
% \end{table}

\subsubsection{Compatibility for ID-based models}
Given the orthogonal nature of our proposed framework to ID-based models, we explore its efficacy and efficiency across a variety of backbones. As shown in Table~\ref{tab:diff_backbone} and Appendix~\ref{appendix:cost}, CELA consistently achieves significant performance improvements and comparable inference and training costs when compared to these backbones, thereby confirming the framework’s model-agnostic nature and extensive compatibility.

\begin{table}
  \caption{Performance comparison of models with different backbones. The language model is RoBERTa.}
  \vspace{-0.3cm}
    \tabcolsep=0.35cm
  \label{tab:diff_backbone}
  \scalebox{0.9}{
  \begin{tabular}{@{}ccccc@{}}
\toprule
\multirow{2}{*}{Model}& \multicolumn{2}{c}{Amazon} & \multicolumn{2}{c}{AppGallery} \\
\cmidrule(r){2-3} \cmidrule(l){4-5}
& AUC & Logloss & AUC & Logloss \\
\midrule
AutoInt                      & 0.8793 & 0.4383 & 0.8426 & 0.4777 \\
$\text{CELA}_\text{AutoInt}$ & \textbf{0.8831} & \textbf{0.4323} & \textbf{0.8470} & \textbf{0.4715} \\
\midrule
DCNv2                       & 0.8809  & 0.4335 & 0.8432 & 0.4771 \\
$\text{CELA}_\text{DCNv2}$  & \textbf{0.8851}  & \textbf{0.4290} & \textbf{0.8472} & \textbf{0.4714} \\
\midrule
DeepFM                      & 0.8798 & 0.4378  & 0.8434 & 0.4771 \\
$\text{CELA}_\text{DeepFM}$ & \textbf{0.8852} & \textbf{0.4263}  & \textbf{0.8483} & \textbf{0.4699} \\
\midrule
DIN                         & 0.8949 & 0.4110  & 0.8437 & 0.4771 \\
$\text{CELA}_\text{DIN}$    & \textbf{0.8996} & \textbf{0.3996}  & \textbf{0.8481} & \textbf{0.4705} \\
\bottomrule

\end{tabular}}
\vspace{-0.2cm}
\end{table}

\subsubsection{Analysis of Alignment w.r.t. Popular Items}
\begin{figure}
    \centering
    \includegraphics[width=\linewidth]{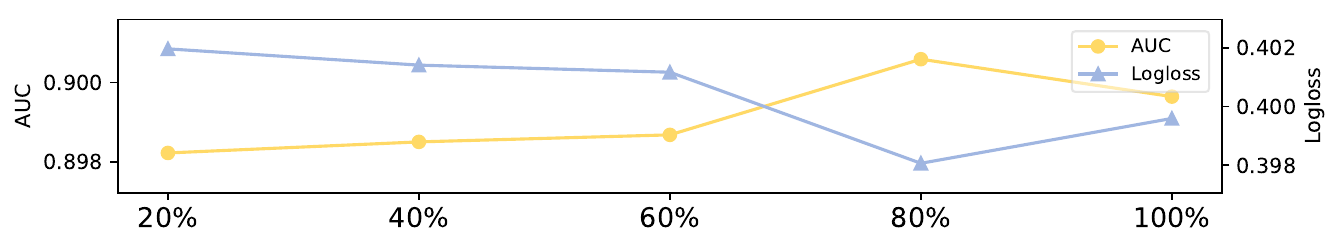}
    \caption{Performance with alignment w.r.t. popular items, evaluated on Amazon. The $x$-axis represents the top percentile of popular items.}
    \label{fig:tophot}
\end{figure}
ID-based models exhibit poor performance in cold-start scenarios, suggesting inadequacies in learning ID embeddings for cold-start items. We aim to ascertain whether aligning poorly learned ID embeddings during the ROMA stage might lead to a decline in performance. To this end, we design an experiment in which CELA's ROMA stage aligns only the top 20\%, 40\%, 60\%, 80\%, and 100\% of items by popularity. Figure~\ref{fig:tophot} indicates that CELA's performance reaches its apex when aligning 80\% of the most popular items. This suggests that limiting the alignment to a subset of popular items facilitates a more accurate mapping of item text representations to the collaborative filtering space while aligning poorly learned ID embeddings will lead to performance degradation.

\subsection{Online A/B Testing}
To evaluate the effect of CELA on real-world scenarios, we deploy CELA and conduct an online A/B test in an industrial app advertising platform, which is one of the most popular app stores.%Huawei AppGallery.

\subsubsection{Deployment}
\begin{figure}
\vspace{-0.3cm}
    \centering
    \includegraphics[width=0.95\linewidth]{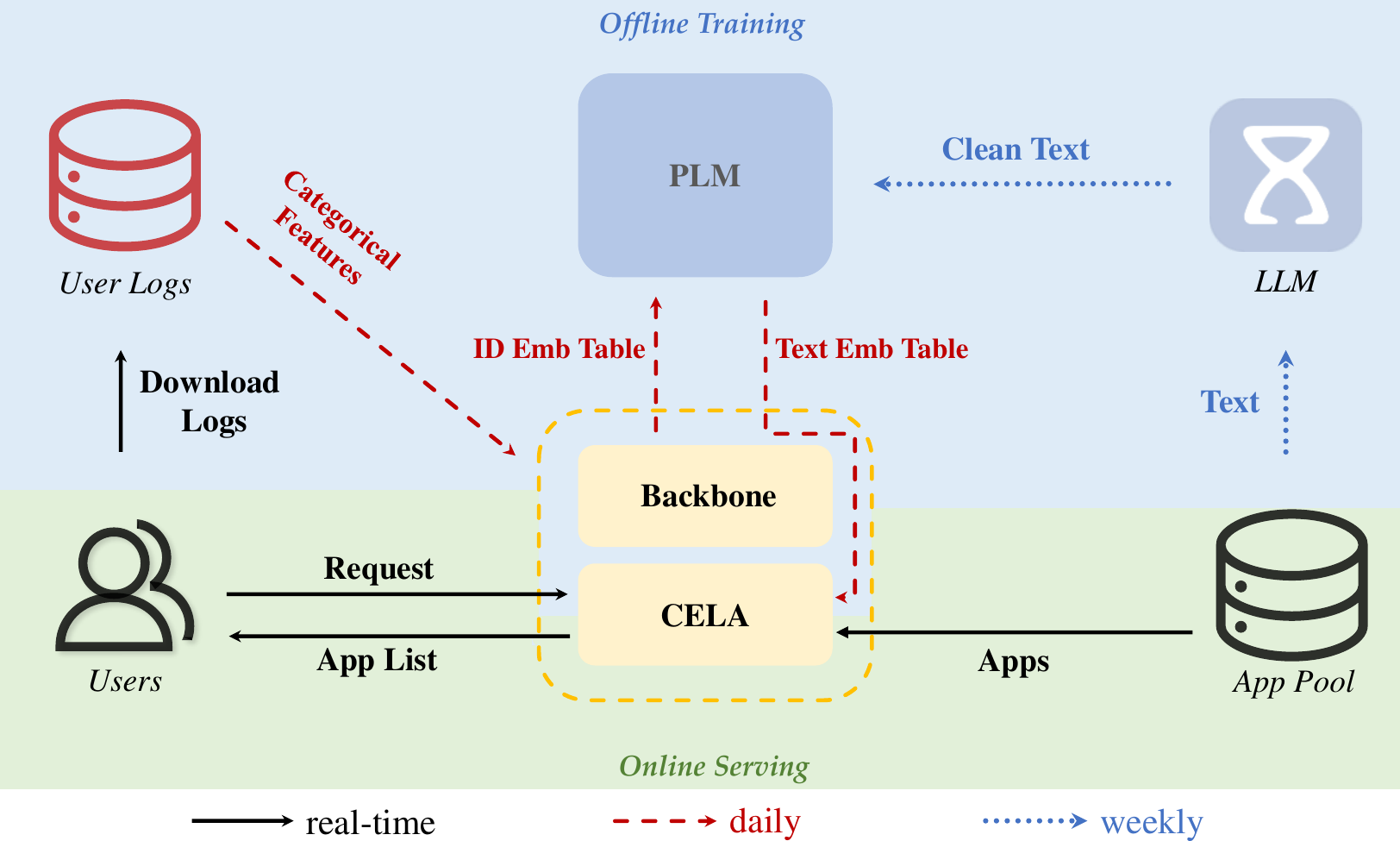}
    \caption{Overview of the app advertising platform.}
    \label{fig:deployment}
    \vspace{-0.2cm}
\end{figure}

The industrial app advertising platform, illustrated in Figure \ref{fig:deployment}, comprises two main components: offline training and online serving.
In offline training, an LLM refines the collected app textual features from the app pool, removing extraneous details such as contact information to enhance conciseness. The clean text is indexed by the app identifier for low-cost storage. Given the infrequent changes in app textual features, training is conducted weekly during the DAP stage. The efficient ROMA stage enables daily alignment of the PLM with ID embeddings from the current ID-based backbone, providing the MF$^2$ stage with the text embedding table and minimizing additional training overhead. Consequently, updates to the MF$^2$ stage also occur daily.
In online serving, when a user initiates a request, the candidate apps are scored by CELA, which retrieves the text embeddings of candidate apps directly from the pre-computed text embedding table, bypassing real-time text processing. The candidate apps are then ranked according to Effective Cost Per Mille (eCPM) and the predicted Download-Through Rate (DTR) scores, with the top-\( k \) apps displayed in the list.

% \begin{itemize}[left=0pt]
%     \item \textbf{Data Preprocessing.} Utilizing a LLM, we refine the collected app descriptions to eliminate extraneous details, such as contact information, thereby enhancing conciseness. The refined text is indexed by the item identifier for low-cost storage.
%     \item \textbf{Update Process.} Given the infrequent changes in app descriptions, weekly training is conducted during the DAP stage. The efficient ROMA stage makes it feasible to align the PLM daily with the ID embeddings from the previous day's ID-based model. This alignment supplies the MF$^2$ stage with the text embedding table, minimizing additional training overhead. Accordingly, updates to the MF$^2$ stage also occur daily.
%     \item \textbf{Online Services.} The online recommendation model retrieves text representations directly from the pre-computed text embedding table via item identifiers, circumventing the need for real-time PLM text processing.
% \end{itemize}

% This deployment facilitates iterative and daily enhancements of both the ROMA and MF$^2$ stages, while the efficient and effective online serving ensures optimal service delivery.

\subsubsection{Online Experimental Results}
\begin{table}
  \caption{Performance of CELA in online A/B test.}
  \vspace{-0.3cm}
  \tabcolsep=0.3cm
  \label{tab:online}
  \scalebox{0.9}{
  \begin{tabular}{cccc}
\toprule
eCPM    & DTR     & Training Time & Inference Time\\
\midrule
+1.48\% & +0.93\% & -19.41\% & +9.69ms \\ %69.73ms\\
\bottomrule
\end{tabular}}
\vspace{-0.2cm}
\end{table}
An online A/B test of CELA is conducted on the advertising platform over three weeks, including tens of millions of daily active users. The treatment group implements CELA's three-stage framework, extending the control group. Each group is trained in a single cluster, where each node contains an 8-core Intel(R) Xeon(R) Gold 6151 CPU (3.00GHz), 32GB RAM, as well as 1 NVIDIA TESLA V100 SXM2 GPU with 32GB memory.

Table~\ref{tab:online} shows the superiorities of CELA: a 1.48\% rise in eCPM and a 0.93\% increase in DTR, indicating a positive influence on user engagement and revenue. Notably, the latency of CELA is below the 100ms threshold
% and compared to the control group's 60.04ms
, which is acceptable for real-world applications.

% we conducted an A/B test of the CELA framework by allocating 1\% of AppGallery's considerable traffic, encompassing tens of millions of daily active users, over a duration of one week. The treatment group applies the three-stage framework of CELA, building upon the foundation established by the control group.

% We focuse on two primary metrics for evaluation: Download-Through Rate (DTR) and Effective Cost Per Mille (eCPM). DTR indicates the effectiveness of app recommendations by measuring download rates post-exposure. A boost in DTR directly enhances user engagement. eCPM, reflecting ad revenue per thousand impressions, gauges the financial efficiency of ads, with improvements signaling increased profitability for developers and the platform.

% Results highlighted in Table~\ref{tab:online} showcase CELA's substantial impact: a 0.92\% rise in Download-Through Rate (DTR) and a 0.97\% increase in Effective Cost Per Mille (eCPM), marking significant advances in user engagement and monetization. Crucially, the operational efficiency of real-time recommendation systems is also gauged by the latency of online serving. CELA adheres to the scenario-standard threshold, maintaining a response time below 100 milliseconds, where the control group is 60.04ms, ensuring CELA's operational viability for deployment in real-world recommendation systems.

\begin{figure}
    \centering
    \includegraphics[width=0.9\linewidth]{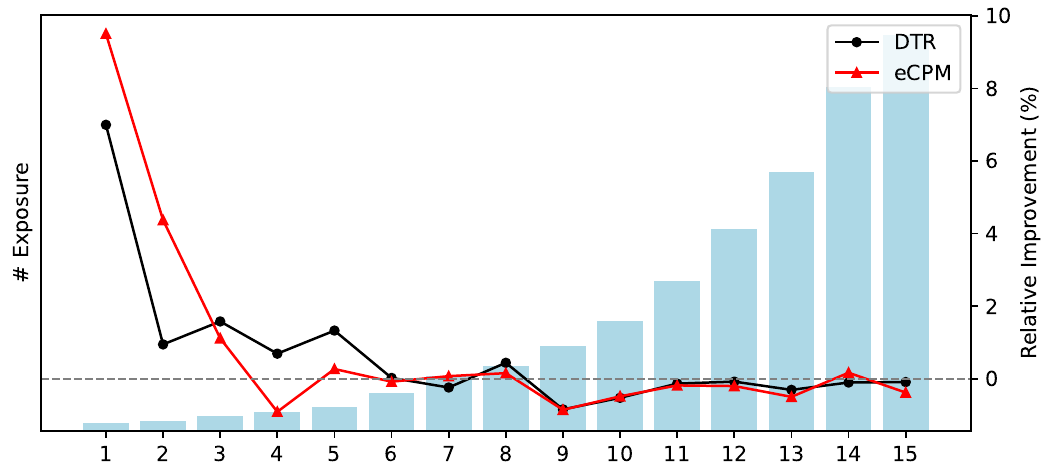}
    \caption{Performance comparison w.r.t. unpopular apps. The bar graph represents the exposure for each group, while the line chart represents the relative improvement compared to the control group.}
    \label{fig:online_cold_start}
    \vspace{-0.3cm}
\end{figure}

To explore the impact of CELA on cold-start items in the real business scenario, we group apps according to baseline exposure and measure the relative improvements in eCPM and DTR. Each group consists of an equal number of apps, focusing on the 15 least popular groups as illustrated in Figure~\ref{fig:online_cold_start}. The results demonstrate that CELA significantly improves performance in unpopular groups. This enhancement is credited to the PLM that leverages external knowledge to create more accurate item profiles.

\section{Conclusion}
In this study, we introduced a novel cost-efficient framework for CTR prediction that leverages the strengths of PLMs and textual features alongside ID-based models. To tailor capabilities of PLMs for recommendation, we implemented domain-adaptive pre-training and item-level alignment, concurrently addressing the inefficiencies associated with PLM training at the interaction level. By abolishing hard prompts and employing PLMs solely as text encoders, we intricately describe user behaviors while preventing token overflow. Our framework is model-agnostic and industry-friendly, with low training overhead and serving latency. Comprehensive experiments conducted on both public and industrial datasets validate CELA's notable enhancements in performance.
\newpage
\bibliographystyle{ACM-Reference-Format}
\bibliography{content/7_ref}

\newpage
\appendix
\section{Corpus Construction}
\label{appendix:corpus}
For illustration, consider the features of an item detailed as follows:
\begin{verbatim}
{
    "item_id": 31852, 
    "brand": "Coxlures", 
    "categories": ["Sports & Outdoors", "Dance"], 
    "description": "The ballet skirt is nice for 
                    dress-up play for ballerinas."
}
\end{verbatim}
Categorical features such as \texttt{item\_id}, \texttt{brand}, and \texttt{categories} can be encoded using conventional methods like one-hot or multi-hot encoding. However, textual features like the \texttt{description}, require language models for encoding. To enhance the language model's understanding of dataset-specific texts, a corpus is constructed from item descriptions. This corpus is organized as follows:
\begin{verbatim}
[
    item 1's description,
    item 2's description,
    ..., 
    item |I|'s description
]
\end{verbatim}

\section{Experimental Settings}
\subsection{Datasets}
Offline experiments are conducted on the datasets below. The statistics are presented in Table~\ref{tab:statistics}.

\begin{table}
  \caption{Statistics of datasets for offline experiments.}
  % \vspace{-0.3cm}
  \label{tab:statistics}
  \tabcolsep=0.2cm
  \scalebox{0.95}{
  \begin{tabular}{lrrrr}
    \toprule
    Dataset&\#Users&\#Items&\#Features&\#Samples\\
    \midrule
    % Toys&207,726&78,099&6&3,234,680\\
    Amazon&300,562&100,682&6&4,494,230\\
    MovieLens&92,911&26,084&4&10,146,329\\
    AppGallery&11,056,819&29,759&8&16,398,193\\
  \bottomrule
\end{tabular}}
\end{table}

\label{appendix:datasets}

\textbf{Amazon Dataset}~\footnote{https://cseweb.ucsd.edu/$\sim$jmcauley/datasets/amazon\_v2/} includes product metadata from the Sports subset on Amazon, which is widely used as a benchmark dataset~\cite{zhou2018deep,zhou2019deep,li2023ctrl}. 
Following \cite{zhou2018deep}, negatives are sampled at a 1:1 ratio. The task involves predicting the likelihood of a user reviewing the next item, using item descriptions as the textual feature and a behavior sequence of up to 20 items.

\textbf{MovieLens Dataset}~\footnote{https://grouplens.org/datasets/movielens/latest/} collects the rating activity from MovieLens~\footnote{http://movielens.org/}. We extract movie summaries as the textual feature from provided hyperlinks and sample 10 million interactions through random sampling of users. The task is to predict whether a user will like the next item, indicated by a rating of 3 or higher, based on a sequence of up to 20 user behaviors.

\textbf{AppGallery Dataset} comprises application exposure and download records sampled from an industrial app advertising platform over eight days with user consent. Data from the first seven days constitute the training set, while data from the eighth day serve as the test set. The task is to predict whether the user will download the next app, using app descriptions as the textual feature and a sequence of up to 14 user behaviors.

% \subsection{Baselines}
\subsection{Implementation Details}
In our offline experiments, models are implemented using \emph{RecStudio}~\cite{lian2023recstudio} and \emph{Transformers}~\cite{wolf-etal-2020-transformers}. We integrate RoBERTa~\cite{liu2019roberta} to enhance the optimal ID-based model. AdamW~\cite{loshchilov2017decoupled} optimizer facilitates all training stages. For the DAP stage, $\tau_1$ ranges from [0.001, 0.01, 0.05,  0.1] and $\alpha$ from [0.1, 0.5, 1.0, 1.5], with a batch size of 32. In the ROMA stage, $\tau_2$ is 1, batch size is 128, the language model's learning rate is 5e-5, and projection units are [128, $f_i d$], with $f_i$ as item-side feature count and $d$ set at 16. The MF$^2$ stage uses a batch size of 1024 and standardized hidden units [128, 64, 64] across ID-based models.
ID-based models that do not focus on user behavior modeling aggregate user behavior sequences using mean pooling.

\section{Experimental Results}
\subsection{Compatibility for ID-based models}
\label{appendix:cost}
We extend the evaluation of CELA’s model-agnostic nature by testing it with a wider range of ID-based models and providing a detailed analysis of the computational costs associated with each integration. The results are presented in Table~\ref{cost}.
It is worth noting that CELA’s training involves both the PLM and the backbone, with the PLM requiring 1.07 hours for pre-training and weekly updates, while the backbone is updated daily.

\begin{table}
    \caption{Training and inference efficiency on AppGallery. The inference time is calculated for one batch and the training time is for the whole process.}
    \vspace{-0.1cm}
    \tabcolsep=0.5cm
    \label{cost}
    \scalebox{0.9}{
    \begin{tabular}{@{}crr@{}}
\toprule
Model                           & Inference (ms)    & Training (h)    \\
\midrule
AutoInt	                        & 62.83	            & 2.90\\
$\text{CELA}_\text{AutoInt}$    & 77.96	            & 6.99\\
\midrule
DCNv2	                        & 59.63	            & 2.28\\
$\text{CELA}_\text{DCNv2}$      & 64.16             & 5.66\\
\midrule
DeepFM                          & 62.67             & 2.88\\
$\text{CELA}_\text{DeepFM}$     & 65.36             & 4.77\\
\midrule
xDeepFM	                        & 60.96             & 2.33\\
$\text{CELA}_\text{xDeepFM}$    & 65.36             & 4.85\\
\midrule
DIN                             & 61.38	            & 1.76\\
$\text{CELA}_\text{DIN}$        & 69.64	            & 5.42\\
\midrule
DIEN	                        & 66.94	            & 2.93\\
$\text{CELA}_\text{DIEN}$       & 75.78             & 5.86\\

\bottomrule
\end{tabular}}
\end{table}

\end{document}